\documentstyle[eqsecnum,aps]{revtex}
\begin{document}
\draft
\title{
An Indication From the Magnitude of CP Violations\\
That Gravitation Is A Possible Cause of Wave-Function Collapse  }     
\author{Daniel I. Fivel}
\address{
 Department of Physics,
University of Maryland\\
 College Park, MD 20742}
\date{October 8, 1997}
\maketitle
\begin{abstract}
We consider experimental evidence for the hypothesis that the
Planck energy, $E_p \approx 10^{19}GeV$, sets the scale $\epsilon$
at which wave function collapse causes deviations from linear
Schr\"{o}dinger evolution. With a few plausible assumptions about
the collapse process,
we first show that the observed CP violation in
$K_L$ decay implies a lower bound on $\epsilon$ 
remarkably close to $E_p$. If the bound is saturated, the
entire CP violation is due to collapse and a prediction made
that the branching ratio for CP violation in the B meson decay will be
$\gamma \approx 10^{-5}$.
 We then show that the assumptions  are
consequences of a simple non-linear, stochastic modification of the 
Schr\"{o}dinger equation with  $\epsilon$ setting the scale of the non-linearity.
\end{abstract}
\vskip.2in\indent
In his {\em Lectures on Gravitation}\cite{FEYNMAN}, R.P. Feynman suggested 
 that we at least consider the
possibility that gravity need not be quantized because it is actually
responsible for  a  deviation from linear
Schr\"{o}dinger evolution at
the Planck scale, $E_p \approx 10^{19} GeV$.
 This is not implausible since
a particle with mass larger than $M_p = E_p/c^2 \approx 10^{-5}$gm has a Compton wavelength smaller
than its Schwarzschild radius.
\vskip.1in 
 Feynman also observed that because the Planck mass
$M_p$ is essentially macroscopic, such a deviation might 
 explain the apparent instability and collapse of Schr\"{o}dinger cat states in
which there is dispersion of a macroscopic observable. Indeed, the measurement problem 
of quantum mechanics arises from the absence of a generalization of the Schr\"{o}dinger equation
that interpolates between linear evolution and collapse. 
 Such a  generalization requires
 a universal energy scale $\epsilon$
to which a dispersion $\Delta$ arising from entanglement can be compared in determining
whether the collapse time $\tau_c$ is large or small in relation to the time scale $\hbar/\Delta$
of linear Schr\"{o}dinger evolution. The question then is whether $E_p$ supplies that scale.
Since there is, as far as
we know, no interaction weaker than gravity, the establishment of a 
{\em lower  bound} for  $\epsilon$ in the vicinity of $E_p$  would lend strong support to the
candidacy of gravity as the cause of collapse. 
\vskip.1in 
The simplest guess we can make from dimensional analysis is
$$ \tau_c = \hbar \epsilon/\Delta^2. \eqno{(1)} $$
If collapse does not actually happen, as some believe, we would have
$\tau_c \to \infty$, i.e.\ $\epsilon \to \infty$. If collapse does happen,
$\epsilon \approx E_p $
 implies that collapse is very slow and difficult to measure. 
\vskip.1in
To obtain a lower bound experimentally one might consider the possibility of detecting collapse by
its effect on the observable EPR correlations of entangled pairs.  But in today's technology such
experiments are only conceivable in the realm of quantum optics where
 $\Delta \approx 1 eV$. Absence of observed collapse would then imply
 $\epsilon >>  1 eV$ . Even if such measurements could be made with hadrons, our bound will only
move up to $1$ GeV, and we are still nineteen  orders of magnitude too low.
\vskip.1in
Fortunately, however, there are entanglement situations involving elementary particles in which 
a conspiracy of circumstances provides an opportunity to detect a much larger lower  bound.
Suppose  we have an entangled state which
is unstable, decays in time $\tau$, and has an observed branching
ratio $\gamma << 1$ for breaking a symmetry. If $\tau_c$ were not very much larger than
$\tau$, there would be time to first collapse into the factorized constituents of
the entangled state. If the branching ratio for violation in such case is $\gamma_c >>
\gamma$, a lower bound on
$\tau_c$ follows. If collapse obeys an exponential law one obtains:
$$ \tau_c \geq \gamma_c \tau/\gamma \quad \Longrightarrow \quad
\left({\epsilon \over \Delta}\right)  \geq \left({\tau\Delta\over
\hbar}\right)\left({\gamma_c\over\gamma}\right),\eqno{(2)}$$
with saturation indicating that all of the violation is coming from the collapse channel.
\vskip.1in
Thus we can obtain a lower bound on $\epsilon$ much larger than $\Delta$ by having
either or both of the following:
(I)  $\tau \Delta/\hbar >> 1 $,
i.e.\ an unusually long lifetime because of the violation of a good conservation law.
(II) $\gamma/\gamma_c$ small, i.e.\ a small observed symmetry breaking but one
that would be large if collapse happens before decay. 
\vskip.1in
\indent
Both of these ingredients are present in the CP violating decay of
the $K_L$ meson which in the quark model is the
state:
$$K_L = \alpha |s,1\rangle|{\bar
d},2\rangle  + \beta |d,1\rangle|{\bar
s,2}\rangle  ,\qquad
\alpha = 1/\sqrt{2}  = - \beta .\eqno{(3)}$$
 This is a superposition of 
  the $K_o$ and ${\bar K}_o$ which are the summands on
the right, but it is an {\em entanglement} of the two-level di-quark system 
indicated by label 1 and the two-level di-antiquark
indicated by label 2. As will be made clear below, the relevant value of the
dispersion $\Delta$ driving collapse is the total of the dispersion in
each of the two entangled systems {\em separately}. Thus it is the dispersion
arising from the mass difference $\delta m  \approx 200$ MeV of the $s$ and $d$ quarks
that we must use for $\Delta$ not the mass difference of the $K_o$ and ${\bar K}_o$
which is zero. It is this large value
together with the very long life of the $K_L,\,$ 
$ \tau \approx 5 \times 10^{-8}s$, resulting from
a combination of strangeness changing and small
phase space that supplies the principal
amplification factor (I). The amplification
factor (II) results from having
$\gamma_c = 0.5$ because the $K_o$ and ${\bar K}_o$ are not CP eigenstates, together with
a small measured CP branching ratio, $\gamma \approx 2 \times 10^{-3}$. Putting these
numbers into (2) we obtain the  remarkable result:
$$
 \epsilon \geq E_p/8\pi. \eqno{(4)}$$
\vskip.1in \indent
We cannot, of course, rule out the possibility that $\epsilon$ is  much larger,
or even infinite. If so this occurrence of a bound comparable to $E_p$ and  
{\em twenty eight orders of magnitude higher than any present experiment could have anticipated}
is just a bizarre coincidence.
 Absent such a coincidence we are forced to conclude that
collapse does indeed  ``happen" and that gravitation is somehow implicated. It also  suggests, but does
{\it not} require, an economical explanation of CP violations, namely  that the bound is saturated, thus
making collapse the sole cause.. This would mean that the CKM phase
is zero. Moreover in the B-system, insertion of the b-d mass difference of 5 GeV for
$\Delta$ and  the lifetime $10^{-12}s$ for $\tau$ then predicts a branching ratio of $\gamma \approx
10^{-5}$.
\vskip.1in \indent
Assuming that
 (4) is indeed a manifestation of a dynamical collapse mechanism, we are motivated
to carry out the task that will occupy the remainder of this paper, namely the
construction of a generalization of Schr\"{o}dinger dynamics with one universal energy parameter
$\epsilon$ whereby one can {\em deduce} the various ingredients of the above argument i.e.\
the validity of (1), the exponential law for collapse, and the independent contribution
to $\Delta$
from the energy dispersions in each constituent of the entangled state. We shall arrive at
the generalization through a sequence of steps intended to show that where choices are available
we have made the simplest consistent with physical constraints.
\vskip.1in \indent
 We know that the modified Schr\"{o}dinger
equation must be non-linear to allow collapse to happen. But we also know \cite{GISIN,PERES,POLCHINSKI}
that non-linearities must
 be stochastic (noisy) or else pathologies will appear, i.e.\ the non-linearity can be exploited
 to send super-luminal signals and reduce the entropy of closed systems.  
\vskip.1in \indent
To keep the notation uncluttered we develop the theory for
two-level entangled states such as (3)  We will assume that the two levels for particle-1 are
$\pm E$ which can be done  by adding a multiple of the unit matrix to the Hamiltonian
matrix. We first develop the equation for the case where only one of the particles has
significant energy dispersion and then extend it to allow contributions from both. We will
thus take the Hamiltonian to be:
$$  H_{tot} = H \otimes I, \qquad
H = E\sigma, \quad \sigma  = \left(\matrix{1 & 0 \cr 0 & -1}\right).\eqno{(5)}$$
\vskip.1in\indent
We obtain a first clue as to the kind of stochastic non-linearity we need by
looking at wave-function collapse thermodynamically. It 
is a peculiar  irreversible process in which an entangled pure state
 such as (3), is driven into another pure state which is factorized. As 
 normally defined all pure states have zero entropy. But one notes that
entangled states contain more information than do pure states. For if a measurement
 is made on particle-1, paying no attention to  particle-2, the
density matrix becomes the reduced density matrix whose entropy is larger, whereas for a factorized
state there would be no change. Thus (3) has  a negative entropy  of entanglement
$$S(x) \equiv x\log x + (1 - x)\log(1 - x), \qquad x = |\alpha|^2, \eqno{(6)}$$
 which rises to zero when collapse takes place. We thus expect
that the non-linear term that we add to the Schr\"{o}dinger equation to induce
collapse should depend on the wave function in such a way that its strength is
correlated with the magnitude of this  entanglement entropy.
\vskip.1in\indent
There is an elegant and  natural way to introduce non-linearities into the
Schr\"{o}dinger equation. If $\Psi$ is an  n-component spin wave function, and $H$ is an $n\times n$
matrix then  
$$\hbar{d\Psi\over dt} = - i H\, \Psi \qquad \Longleftrightarrow \qquad
\hbar {d\Psi\over dt} = -iH{\partial\over \partial \Psi^*}|\Psi|^2, \qquad |\Psi|^2 = \sum_j|\Psi_j|^2.
\eqno{(7)}$$
This is the so-called ``geometric formulation of quantum mechanics"\cite{KIBBLE,HESLOT} which makes
the Schr\"{o}dinger equation exhibit the (K\"{a}hler)
symplectic structure\cite{HUGHSTONA,ASHTEKAR} of Hamilton's
equations for conjugate dynamical variables
$\Psi$ and $\Psi^*$.
One may now look for suitable modifications of the Schr\"{o}dinger equation by adding something
to $|\Psi|^2$ which is more than quadratic in the $\Psi$'s and  is thus like adding anharmonic terms
to a classical oscillator. Our problem is to find such a quantity that is correlated with the entanglement
entropy and is expressed simply in terms of the $\Psi$'s. 
\vskip.1in
We obtain a second clue as to the kind of quantity we need from the observation that entanglement is
invariant under one-particle unitary transformations, i.e.\
$$|\Psi\rangle = \sum_{j,k }\psi_{jk}|j,1\rangle |k,2\rangle \quad
\to\qquad  (U \otimes V)|\Psi\rangle . \eqno{(8)}$$
We can construct  invariants under such transformations by observing that there
is a natural map from $|\Psi\rangle$ to the operator
$$\psi = \sum_{j,k }\psi_{jk}|j,1\rangle \langle k,2| ,\eqno{(9)}$$
which is mapped under $U\otimes V$ to $U\psi V^{\dagger}$. Thus functions of
the traces of powers of $\psi^{\dagger}\psi$ are invariants as is 
the determinant
$${\cal E}(\psi) \equiv \hbox{Det }(\psi^\dagger\psi).\eqno(10)$$
For the situation in which we are interested, $\psi$ is a two by two matrix so  that all
of the invariants are functions of  $\hbox{Tr }(\psi^\dagger\psi)$ 
 and  ${\cal E}$.  Since $\hbox{Tr }(\psi^\dagger\psi)$ is just the
norm $|\Psi|^2$ in (7) which gives the linear Schr\"{o}dinger equation, we are led
to consider whether ${\cal E}$ might be the quantity  we are seeking.
\vskip.1in\indent
In 
the case of the state (3) we have
${\cal E} = |\alpha|^2|\beta^2| = x(1 - x)$ so that it is correlated with
 the entanglement entropy (6) in the right way, having its maximum when the
$|S|$ is largest and falling to zero when $|S| = 0$. That it is a quite
general measure of entanglement can be seen from the following argument:
What characterizes entanglement is the inability to make a
prediction about the behavior of one constituent independently of its partner. Thus in a
completely entangled state  a constituent can be found with equal likelihood in any state if
no measurement is made on its partner. In contrast, for a system in the factorized state
$|1\rangle|2\rangle$, it is certain that particle -1 will not be found in a state 
orthogonal to
$|1\rangle$. Now one sees that a necessary and
sufficient condition for there to be zero probability of finding a particle in 
some state when no measurement is made on the other, is that such a state be an
eigenstate of $\psi^\dagger\psi$ with eigenvalue zero. 
There is an eigenstate with
eigenvalue zero if  and only if ${\cal E}$ vanishes.
\vskip.1in\indent
It appears then that we should be able to build our modified dynamics by rewriting (7)
in terms of $\psi$ and replacing $|\Psi|^2$ by a suitable combination of $\hbox{Tr }(\psi^\dagger\psi)$
and ${\cal E}$. 
Taking advantage of the identity
$$
\hbox{Det} (I + \nu A) \equiv I + \nu \hbox{Tr}(A) + \nu^2 \hbox{Det}
(A),
\eqno{(11)}$$
valid for 
any $2 \times 2$ matrix $A$ and scalar $\nu$,
we are led to consider an equation of the form
$$ {\hbar\over\epsilon}{d \psi \over dt} = \pm\sigma{\partial\over\partial_{\psi^*}}
\hbox{Det }(1 \mp i \eta
\psi^\dagger\psi),\qquad \eta = E/\epsilon,\eqno{(12)}$$
in which we have one new universal  parameter $\epsilon$. 
Expanding the determinant by (11) and working out the partial derivatives one obtains:
$$
\hbar {d\psi\over dt} =  -iH\psi \pm \eta \hbox{Det }(\psi^\dagger\psi) 
H \psi^{\dagger -1} ,\eqno{(13)}$$
which shows that the equation reduces to the ordinary Schr\"{o}dinger equation plus
a nonlinear term that disappears if we let $\epsilon \to \infty$. This would be the
case if collapse did not really happen at all.
\vskip.1in
 The linear
Schr\"{o}dinger equation has a time reversal symmetry that shows up in its invariance under $t \to -
t$ and $\psi \to \psi^*$. The non linear term which has no
$i$ does not preserve this symmetry. With a random  sign  in (13) the symmetry can be preserved as
an ensemble average. This suggests a simple and natural way to introduce the noise:
{\em We assume that
during the evolution of the state of a particle,  the signs in (13)
fluctuate randomly.}
\vskip.1in
The description of the noise is not complete, of course, until we also specify the
intervals between the random sign fluctuations.
 Before doing so
let us  
 examine the solution of the equation for a {\em fixed} choice of sign. Because of the invariance
of the determinant, the linear term in (13) is transformed away by going to the interaction
picture, i.e.\ by the trasnformation $\psi \to e^{-iH t} \psi$. The
$\psi$-matrix representing the state $(3)$ will be initially diagonal and, because of the
absence of coupling, will remain diagonal. We thus arrive at a pair of coupled differential
equations for $x =
|\alpha|^2$ and
$y = |\beta|^2$ , namely:
$${dx\over dt} = - {x y\over \tau_c},\qquad {dy\over dt} = {  x y \over \tau_c}\eqno{(14)}$$
with $\tau_c$ given by (1), and $\Delta = E \sqrt{2}.$
One observes that $x + y$ is conserved, and the process {\em terminates} when either $x$ or $y$ vansishes.
The solution for $x$ is seen to be:
$ x = 1/(1 + e^{t/\tau_c})$ which is essentially exponential with characteristic time $\tau_c$.
\vskip.1in
We now introduce the noise in the manner suggested above by letting the signs in (13) fluctuate
randomly. 
 Writing $dx = \pm \delta, \, dy = \mp \delta$ with $\delta  =xy dt/\tau_c$ one sees
that if the fluctuations occur in such a way that $\delta$ is held fixed the process is 
a random walk with boundary, sometimes called the ``gambler's ruin"\cite{PEARLE,FELLER}. One thinks
of
$x$ and
$y$ as the fortunes of two gamblers who bet $\delta$ on each toss of a coin, the victor being
indicated by the sign. The game ends when one is wiped out. This game is known to be a {\em
martingale} i.e.\ the probability of winning is proportional to ones' initial fortune. Hence the
probability for $x \to 1$ is $x$ and for $y \to 1$ is $y$ which is just the quantum mechanical
collapse rule {\em but now produced dynamically rather than as a separate postulate}.
\vskip.1in
However, this stochastic process is too noisy and will result in an infinitely long process
when we let $dt$ go to zero. For if the stake gets small the weaker player can have runs of
luck so that the average length of the game increases without limit. The way around this
is to let the ``game" be played as the game of ``double or nothing"\cite{FIVEL} in which the weaker
of the two players bets the full amount of his fortune on any play. This game is also a
martingale\cite{MAITRA}, but its average length  is two (or 1 if $ x = y = 1/2$). 
\vskip.1in
As noted earlier we only considered the effect of the energy dispersion arising
from one of the two particles in the entangled state. Effects of the second particle
are included by adding a second term on the right of (12) identical to the first but with
 $\sigma$ acting from the right on the gradient and the E value
appropriate to the second particle. In general there need be no correlation between
the signs of the noise in the two terms, and the result will be to simply add the
dispersions. The fact that it is the sum of the energy dispersions that is relevant rather than the
dispersion of the total energy (which is zero) is easily seen physically by the example of
an entangled state consisting of  a live cat A and dead cat B with  dead cat A and live cat B.
 There is no dispersion of the  total temperature, but the state must clearly collapse
rapdily due to the elevated free energy associated with the dispersion of the temperatures in each
cat separately.
\vskip.1in
We have now produced a stochastic, non-linear modification
of the Schr\"{o}dinger equation that interpolates between linear deterministic evolution and
non-linear, stochastic collapse and from which we could derive
 (1) and the exponential collapse law with no stochastic tails..
Our final task is to show that our modified Schr\"{o}dinger equation enjoys the same 
 ``peaceful coexistence" with the other basic
laws of physics that the linear Schr\"{o}dinger equation enjoys. This is particularly important
because we know\cite{GISIN,PERES,POLCHINSKI} that the noise is essential for this.
\vskip.1in
(1) The second law of thermodynamics: 
Consider the change in the average entanglement of an ensemble with a given $x$ resulting from
a noise fluctuation. We have
$$0.5(x+\delta)(1-x - \delta) + (x - \delta)(1 - x + \delta) - x(x - 1) = - \, \delta^2\eqno{(15)}$$
Thus the average entanglement always decreases which means that the entropy always increases.
Since the maximum average entanglement decrease is achieved for the largest  $\delta$
consistent with the  given $x$ which is the minimum of $x$ and $1 - x$ we see that {\em 
the ``double or nothing" noise rule is equivalent to the requirement that entropy increase at the
maximum possible rate}.
\vskip.1in
(2) Conservation of energy: One may calculate 
${d\over dt}\hbox{Tr }(\psi^\dagger H \psi)$ from (13). Noting that the total time for the
process is $\approx 2\tau_c$ one obtains an upper bound on the energy fluctuation. Comparing
this with the dispersion, one finds that the fluctuation lies within the uncertainty principle
limit.
\vskip.1in
(3) Causality: The random sign fluctuations reproduce the collapse postulate
of conventional quantum mechanics, and so the outcome of measurements has the same
unpredictability. But it is this unpredictability that prevents capitalizing on EPR
correlations to send super-luminal messages. Thus the status of the non-linear, stochastic
theory vis a vis causality is the same as that of the linear equation plus collapse postulate.
There are grounds for optimism that one can develop an explicitly covariant version
because of the  invariance  of  $\hbox{Det }(\psi^\dagger \psi)$ under all possible one-particle
unitary transformations.
\vskip.1in
Since we have  proposed a resolution of the measurement paradox, let us
point out its salient differences from other proposed resolutions{\cite{ALBERT,LEGGETT,SHIMONY}. As
in other dynamical theories, collapse actually happens, rather than merely appearing to happen. In
this respect it differs from relative state and modal interpretations.\cite{BUB} 
The noise is internal
so it also happens in closed systems, a feature missing in the so-called\cite{BELL} FAPP  ("for all
practical purposes") approaches which rely on external noise reservoirs to produce
decoherence. Finally it differs from spontaneous localization
theories by denying favoritism to any direction in Hilbert space, but rather relying on 
the entanglement
and thermodynamics to decide the direction of collapse by lowering the free energy. Having noise
fluctuations occur at intervals influenced by the stochastic process itself 
 we  avoid the necessity of having external ``hits"\cite{GHIRARDI} and the necessity of having more
than one fundamental parameter. If $\epsilon$ is simply related to $E_p$ as the bound (4)
suggests, there will be {\em no} new parameters in our modified equation.
\vskip.1in 
The theory presented above is restricted to the simplest dynamical situation, that of two
systems, each with two levels. Since all measurements in principle can be reduced to yes-no questions,
this treatment is adequate for demonstrating how the measurement paradox may be resolved. 
However, to make it into a complete theory we must extend the non-linear equation to systems
of many particles and to infinite dimensional Hilbert spaces. We anticipate that the thermodynamic
view of collapse will be extremely useful for this, because the optimal decreasing of entanglement
free energy becomes a physical basis by which the system ``decides" along
which basis in Hilbert space to collapse.
\vskip.1in
When nearing the conclusion of this paper the work of L. Hughston\cite{HUGHSTONB} was brought to my
 attention. He has clearly recognized that (1) will be obtained in stochastic evolution, and
his non-linear stochastic equation also has the symplectic geometric form.
In considering the possibility that the Planck scale may be involved he notes
the relative accessibility of
the combination  $\hbar E_p
\approx 8 MeV^2 s$.
 His analysis of the geometric content of stochastic dynamics 
will no doubt be invaluable in seeking to generalize the theory to more complex systems. It might also
provide machinery for attacking the puzzle of why gravitation is playing the role it seems to
be playing  in the dynamics of wave function collapse. 
\vskip.1in
{\em Acknowledgements} I wish to thank  A.\ Dragt, A.\ Jawahery, A.\ Shimony, G.\ Snow,
and J.\  Sucher for  for useful discussions.
I also owe a profound debt of gratitude  to  T.\ Jacobson for much insightful criticim at
various stages of this work, and most importantly for pointing out the relationship of the
entanglement measure to the relative entropy.

{~}

{~}

\end{document}